# Controllable Schottky Barriers between MoS$_2$ and Permalloy

*Weiyi Wang, Yanwen Liu, Lei Tang, Yibo Jin, Tongtong Zhao and Faxian Xiu\**

State Key Laboratory of Surface Physics and Department of Physics, Fudan University, Shanghai 200433, China
E-mail: faxian@fudan.edu.cn

**Abstract**

**MoS$_2$ is a layered two-dimensional material with strong spin-orbit coupling and long spin lifetime, which is promising for electronic and spintronic applications. However, because of its large band gap and small electron affinity, a considerable Schottky barrier exists between MoS$_2$ and contact metal, hindering the further study of spin transport and spin injection in MoS$_2$. Although substantial progress has been made in improving device performance, the existence of metal-semiconductor Schottky barrier has not yet been fully understood. Here, we investigate permalloy (Py) contacts to both multilayer and monolayer MoS$_2$. Ohmic contact is developed between multilayer MoS$_2$ and Py electrodes with a negative Schottky barrier, which yields a high field-effect mobility exceeding 55 cm$^2$V$^{-1}$s$^{-1}$ at low temperature. Further, by applying back gate voltage and inserting different thickness of Al$_2$O$_3$ layer between the metal and monolayer MoS$_2$, we have achieved a good tunability of the Schottky barrier height (down to zero). These results are important in improving the performance of MoS$_2$ transistor devices; and it may pave the way to realize spin transport and spin injection in MoS$_2$.**

Molybdenum disulphide (MoS$_2$) is a layered two-dimensional (2D) material, which draws intensive attention because of its unique electrical, optical and mechanical properties. It has been considered as a promising candidate for various future nanoelectronic and spintronic applications. It is a semiconductor with an indirect bandgap (1.2 eV) for its bulk material and a direct bandgap (1.8 eV) for monolayer MoS$_2$.[1-6] The presence of the direct bandgap in monolayer MoS$_2$ allows room-temperature FETs with an on/off current ratio exceeding $10^8$.[7] Studies also point out that MoS$_2$ could be used in sensors[8] or for photoluminescence[9] applications. The spin transport properties were theoretically predicted with a spin diffusion length of 400 nm at room temperature, and it becomes longer than 800 nm at low temperature ($T < 77$ K).[10] Because of the semiconducting nature, the contacts between MoS$_2$ and metal electrodes naturally forms Schottky barriers, which plays a crucial role in spin transport. To cope with this problem, theorists proposed that titanium is a suitable electrode material because of favorable geometry and large electronic density of state at the interface.[11] And experimentalists found that devices with scandium electrodes have very good performance because of low metal work function.[12] However, until now there is not much research on the ferromagnetic contact with MoS$_2$,[13,14] which may lead to the reduced conductance mismatch and the enhanced spin polarization.

In this letter, we study the contact between MoS$_2$ and ferromagnetic permalloy (Py) electrodes. The Schottky barrier height (SBH) is extracted by 2D thermionic emission theory.[13,15] For multilayer MoS$_2$, we observed a barrier height of -5.7 mV. This negative Schottky barrier leads to a perfect Ohmic contact between MoS$_2$ and Py electrodes which dramatically enhances the field effect transistor (FET) performance of MoS$_2$: the mobility of multilayer MoS$_2$ with a single back gate reaches 55 cm$^2$ V$^{-1}$s$^{-1}$ at low temperature. For monolayer MoS$_2$, the SBH has a positive value, which can be readily tuned by both back gate voltage and by the insertion of an Al$_2$O$_3$ tunneling layer. When inserting 2.5 nm Al$_2$O$_3$, the SBH is reduced from 80.2 to 2.7 mV; by applying a back gate voltage of 30 V on the samples without Al$_2$O$_3$, the SBH is reduced to -8.3 mV. These two approaches of tuning the barrier height are important for realizing spin injection into monolayer MoS$_2$.

## Results

**Structure of the FET devices and method for extracting SBH.** A schematic diagram of our devices is shown in **Figure 1a**. Electrode on Si substrate is used as a back gate to tune Fermi level and carrier density. **Figure 1b** shows a scanning electron microscopy (SEM) picture of the device. The channel length of the device is 2 μm and the width of exfoliated $MoS_2$ is about 4 μm. The channel material is verified to be tri-layer $MoS_2$ by a Raman spectrum[16] as shown in **Figure 1c**. The devices were characterized by two-probe *I-V* measurement for Schottky barrier height and by $I_D$-$V_G$ measurement for mobility.

The forward *I-V* characteristics of an ideal Schottky diode can be described as follows:[17]

$$I_f = I_s \exp(qV_f/k_B T) \qquad (1)$$

where

$$I_s = AST^2 \exp(-q\Phi_B/k_B T) \qquad (2)$$

$I_s$ is the diode saturation current, $A$ is the Richardson constant, $S$ is the contact area of junction, $q$ is the electron charge, $\Phi_B$ is the Schottky barrier height, and $k_B$ is the Boltzmann constant. As the device is thin enough to be treated as a 2D material, the drain-source current $I_{DS}$ can be defined by 2D thermionic emission equation,[18] which employs the reduced power law $T^{3/2}$ for a 2D transport channel:

$$I_{DS} = A^*_{2D} ST^{3/2} \exp\left[-\frac{q}{k_B T}\left(\Phi_B - \frac{V_{DS}}{n}\right)\right] \qquad (3)$$

where $A^*_{2D}$ is the 2D equivalent Richardson constant, $n$ is the ideality factor, and $V_{DS}$ is the drain-source bias voltage. To determine the Schottky barrier height $\Phi_B$, temperature dependent *I-V* measurements were carried out. **Figure 2a** shows the *I-V* curves at several temperatures on a logarithmic scale. To investigate the barrier, it is common to use Arrhenius plot, *i.e.*, $\ln(I_{DS}/T^{3/2})$ against $1000/T$ for various $V_{DS}$ in **Figure 2b**. By fitting the data to each $V_{DS}$, we obtained the slopes with $S = -\frac{q}{1000k_B}\left(\Phi_B - \frac{V_{DS}}{n}\right)$. Then by plotting the slopes as a function of $V_{DS}$, the SBH could be extracted from the y-intercept $S_0 = -\frac{q\Phi_B}{1000k_B}$ (**Figure 2c**).

**Schottky barrier height between Py and tri-layer MoS₂.** For our tri-layer MoS₂ transistor device, $\Phi_B$ is found to be -5.7 mV in the temperature regime of 100 ~ 200 K. Such a negative Schottky barrier produces a good Ohmic contact with a perfect linear *I-V* curve between MoS₂ and Py electrodes (**Figure S1**). Similar results were also reported in *p*-type MoS₂ transistors with MoO$_x$ electrodes.[19] According to the Schottky-Mott model[20], we could roughly estimate the SBH based on the work function of the metal $\Phi_{metal}$ relative to the electron affinity (or vacuum ionization energy) of the semiconductor $\chi_{semi}$:

$$\Phi_B \approx \Phi_{metal} - \chi_{semi} \quad (4)$$

The negative Schottky barrier suggests that the work function of Py is slightly smaller than the affinity of MoS₂, as shown in **Figure 1d**.

To investigate the FET performance, we have performed temperature-dependent $I_D$-$V_G$ measurements. **Figure 3a** shows typical gate-dependent conductance curves at different temperatures and **Figure 3b** displays the temperature-dependent conductance under different gate voltage. It is noted that when applying a small back gate voltage ($V_G$ < 35 V), the channel tri-layer MoS₂ shows an insulating behavior that the conductance decreases as temperature decreases. While $V_G$ >35 V the conductance increases as temperature decreases. This is a hallmark of metallic behavior which suggests that the tri-layer MoS₂ has entered a metallic state.

The field effect mobility can be calculated from the linear regime (40~50 V) of conductance curves using the following expression[21]:

$$\mu_{FET} = [dG/dV_G] \times [L/(WC_{ox})] \quad (5)$$

where $dG/dV_G$ is the slope of the conductance curve in the linear regime, $L = 2$ μm is the channel length, $W = 4$ μm is channel width and $C_{ox} = 1.3 \times 10^{-4}$ F m$^{-2}$ is the capacitance between the channel and the back gate per unit area ($C_{ox} = \varepsilon_0 \varepsilon_r/d$; $\varepsilon_r = 3.9$; $d = 270$ nm). **Figure 3c** shows the extracted mobility as a function of temperature under different drain-source voltage. The mobility is nearly independent of temperature when $T < 20$ K, indicating that the scattering of charged impurities is reduced by drain-source voltage.[22,23] At higher temperatures ($T > 100$ K), the mobility of the tri-layer MoS$_2$ is mainly influenced by the phonon scattering.[23,24] Fitting to the expression $\mu \sim T^{-\gamma}$, the range of exponent part $\gamma$ can be obtained between 0.47 and 0.66. This value is much smaller than the theoretical prediction, *i.e.*, $\gamma = 1.69$ for single-layer MoS$_2$[25] or $\gamma = 2.6$ for bulk crystals[26], indicative of a weak electron-phonon interaction.[20] The mobility of our tri-layer MoS$_2$ exceeds 55 cm$^2$V$^{-1}$s$^{-1}$, which is a comparatively high value for single back gate MoS$_2$ transistor devices. As the Schottky barrier can significantly impact the electron mobility,[12] the high mobility of our device also provides a strong evidence of a low SBH between Py and tri-layer MoS$_2$.

**Schottky barrier height between Py and monolayer MoS$_2$.** Different from the tri-layer, the SBH between Py and the monolayer MoS$_2$ turns out to be 80.2 mV (**Figure 4d**). This can be understood by the fact that the monolayer MoS$_2$ has a large bandgap of 1.8 eV and consequently it has a smaller electron affinity.[4] Thus, the increase of Schottky barrier height in the monolayer system is consistent with the Schottky-Mott model (Equation 4), *i.e.*, for the same metal work function, when the vacuum affinity decreases, the Schottky barrier height increases.

However, great caution must be exercised when the Arrhenius plot is used to extract SBH. The data in **Figure 4Figure 4a** are not completely linear for the entire temperature range. In the high temperature regime ($T > 130$ K), the data show negative slopes, corresponding to the positive SBH. At low temperatures, however, $\ln(I_{DS}/T^{3/2})$ versus $1000/T$ has positive correlations, suggesting negative SBH. These fitted results are contradictory to the observation of "S" shape *I-V* curves at low temperatures (**Figure S2**), which are strong evidence of positive Schottky barriers. To explain this contradiction, one needs to take the semiconducting nature of MoS$_2$ into consideration. As the temperature decreases, the resistivity of monolayer MoS$_2$ increases and the device reaches an "off" state. It is well known that at the off state the channel resistance is too large that the current does not change much with temperature.[20] Thus, the thermionic emission equation is no longer suitable to describe the current. With this limitation, the SBH could only be extracted in the high temperature regime (above the turning points) using 2D thermionic emission equation, while the "V" shaped turning point in the Arrhenius plot represents the entrance of device "off" state.

To investigate the tunability of the Schottky barrier, we have performed the gate-dependent *I-V* measurements (**Figure 4a-c**). When $V_G = 0$ V, the turning point is about 130 K. As the voltage is increased to 10 and 20 V, the turning point shifts to 20 and 8 K, respectively. Further increasing the gate voltage (up to 30 V) makes the turning point completely vanished (**Figure S3**). For traditional field effect transistors, the channel conductance can be tuned by gate voltage. When $V_G > 0$ V, the conduction channel is broadened such that it needs a much lower temperature to turn the device off, which explains the systematic shift of the turning point. **Figure 4d** summarizes the SBH as a function of $V_G$: the SBH is reduced to 20.4, 1.1 and -8.3 mV when $V_G$ changes from 10, 20 to 30 V, respectively. This is attributed to the upwards shift of Py Fermi level by positive gate voltage (**Figure 4d inset**).[27]

For the spin injection into monolayer MoS$_2$, the contact between ferromagnetic metal and semiconductor is crucial.[28] The spin polarization of injected carriers through Ohmic contact is extremely small due to the conductance mismatch.[13] In order to alleviate this issue, the resistance-area (RA) product should be designed properly to obtain a significant spin polarization and magnetoresistance.[29] In conventional 3D semiconductors, the Schottky barrier can provide the required resistance,[30] but the surface doping is needed to facilitate single step tunneling.[31] Another approach to obtain the proper RA product is to achieve a pinning-free FM/Oxide/SC interface by inserting a tunneling oxide layer, which could lower the SBH and the resistance could be tuned by the thickness of the oxide layer. Here we choose $Al_2O_3$ as the tunneling oxide layer, which is well known for high spin injection efficiency and large tunneling magnetoresistance in magnetic tunnel junctions.[32-37]

Similar to the back gate voltage, the insertion of $Al_2O_3$ layer also causes the "V" shaped turning point in the Arrhenius plot to shift towards lower temperatures (**Figure 5a-c**). When the thickness of Al is 0.8, 1.7 and 2.5 nm, the turning point shifts to 40, 25 and 10 K respectively. This can be explained by the reduction of conductance mismatch[38] (or reduced SBH) between the electrodes and monolayer $MoS_2$ as $Al_2O_3$ thickness increases. For that reason, the shift of the turning point could be regarded as an evidence of the alleviation of conductance mismatch with the inserted $Al_2O_3$ layer. The corresponding SBH is extracted to be 32.1, 15.9 and 2.7 mV (**Figure 5d**), respectively. Such a dramatic decrease shows an effective control of the SBH via changing the thickness of oxide layer. By the combination of applying a gate voltage and inserting an $Al_2O_3$ layer, we could minimize the conductance mismatch and tune the Schottky barrier height down to zero, which may help to achieve the proper RA product.

## Discussion

In conclusion, we have investigated the properties of $MoS_2$ FET with ferromagnetic Py electrodes; the Schottky barrier height is extracted using 2D thermionic emission analysis of *I-V* curves. For the tri-layer $MoS_2$, there is a negative Schottky barrier between Py and $MoS_2$ and this Ohmic contact yields a high mobility due to low contact resistance. For the monolayer

MoS$_2$, there is a positive Schottky barrier, which is dramatically reduced either by applying a gate voltage or inserting a tunneling Al$_2$O$_3$ layer. To some extent, the insertion of Al$_2$O$_3$ layer also alleviates the conductance mismatch. Such control of Schottky barrier paves the way of proper design of the RA product, which sheds light on the future research of spin transport and spin injections in MoS$_2$.

**Experimental Section**

Multilayer MoS$_2$ is obtained through mechanical exfoliation from bulk MoS$_2$ crystals onto pre-patterned SiO$_2$/Si substrate (the thickness of SiO$_2$ is 270 nm). FET devices were fabricated by *e*-beam lithography (EBL) using PMMA/MMA bilayer polymer. Subsequently, Py electrodes are deposited by magnetron sputtering, followed by a deposition of gold layer to protect Py from oxidation. All the data are measured from electrode 1 and 2 in **Figure 1b**. For the monolayer transistors, MoS$_2$ is obtained via chemical vapor deposition (CVD) using high purity molybdenum and sulphur as the source materials, similar to the previous report.[39] After growth, they were transferred onto clean SiO$_2$/Si substrate for the following EBL process using PMMA stamping method[40]. The tunneling Al$_2$O$_3$ layer was produced as follows: first a thin layer of Al was deposited by *e*-beam evaporation (the SEM picture in **Figure S4** shows a high-quality Al layer without visible pinholes). Then the samples were placed in the air overnight for natural oxidation to develop Al$_2$O$_3$. The thickness of Al$_2$O$_3$ can be estimated by $d_{Al_2O_3} \approx 1.66\ d_{Al}$.[41] Before measurement, the devices were annealed at 360 K for two hours in vacuum to remove polymer residues between the interface of Py and MoS$_2$.[42,43]

# Figure 1

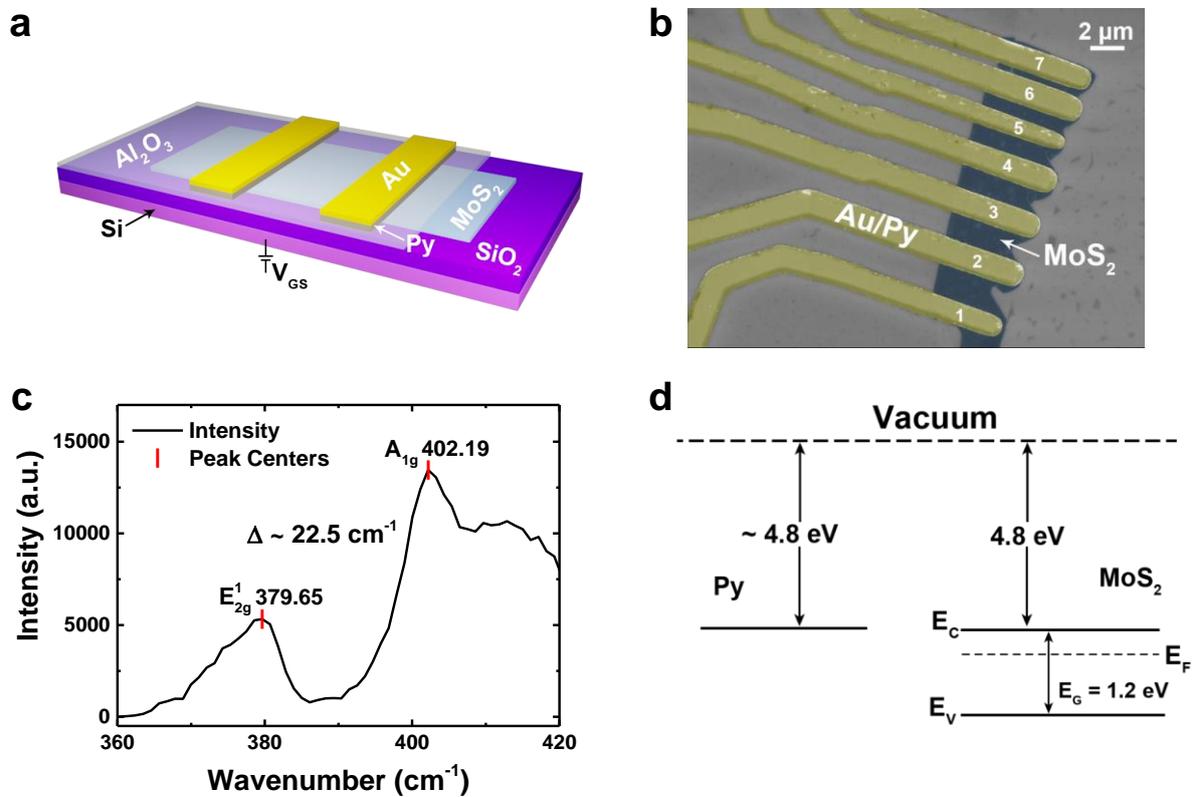

**Figure 1.** (a) A schematic diagram of the device. Py electrodes are covered by Au to prevent oxidation. (b) A SEM image of the multilayer MoS$_2$ device. All the data are measured from electrode 1 and 2. (c) A Raman spectrum of the multilayer MoS$_2$. It is estimated to be a tri-layer MoS$_2$ through the distance of ~22.5 cm$^{-1}$ between two vibrating modes (in-plane mode $E_{2g}^{l}$ and out-of-plane mode $A_{1g}$).[16] (d) A band diagram of Py and MoS$_2$. $E_F$ of Py is close to $E_C$ of the tri-layer MoS$_2$.

# Figure 2

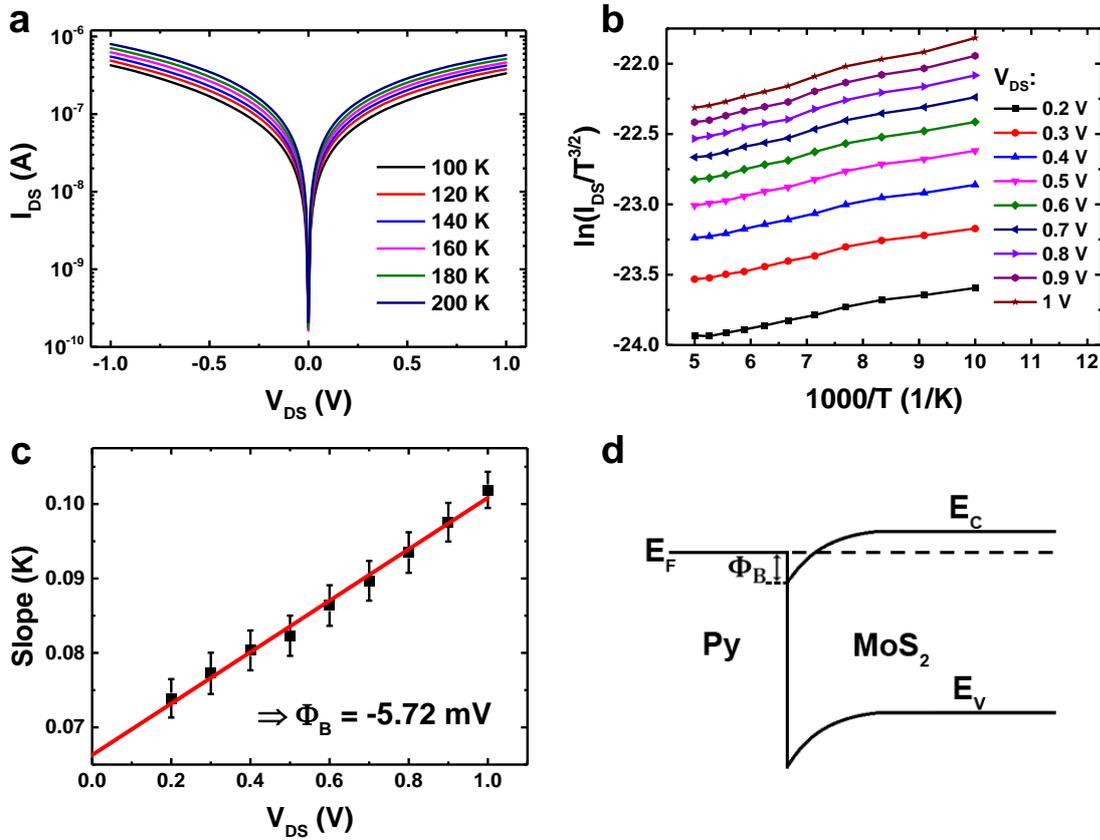

**Figure 2.** (a) *I-V* curves of the tri-layer MoS$_2$ FET device with Py directly contacting to MoS$_2$ from *T* = 100 to 200 K. (b) Arrhenius plot $\ln(I_{DS}/T^{3/2})$ vs $1000/T$ at different drain-source voltages ($V_{DS}$). (c) Extraction of $\Phi_B$ via the y-intercept value. Each data point here represents the slope obtained from the Arrhenius plot in (b) under a specific $V_{DS}$. (d) A band diagram showing a negative Schottky barrier.

**Figure 3**

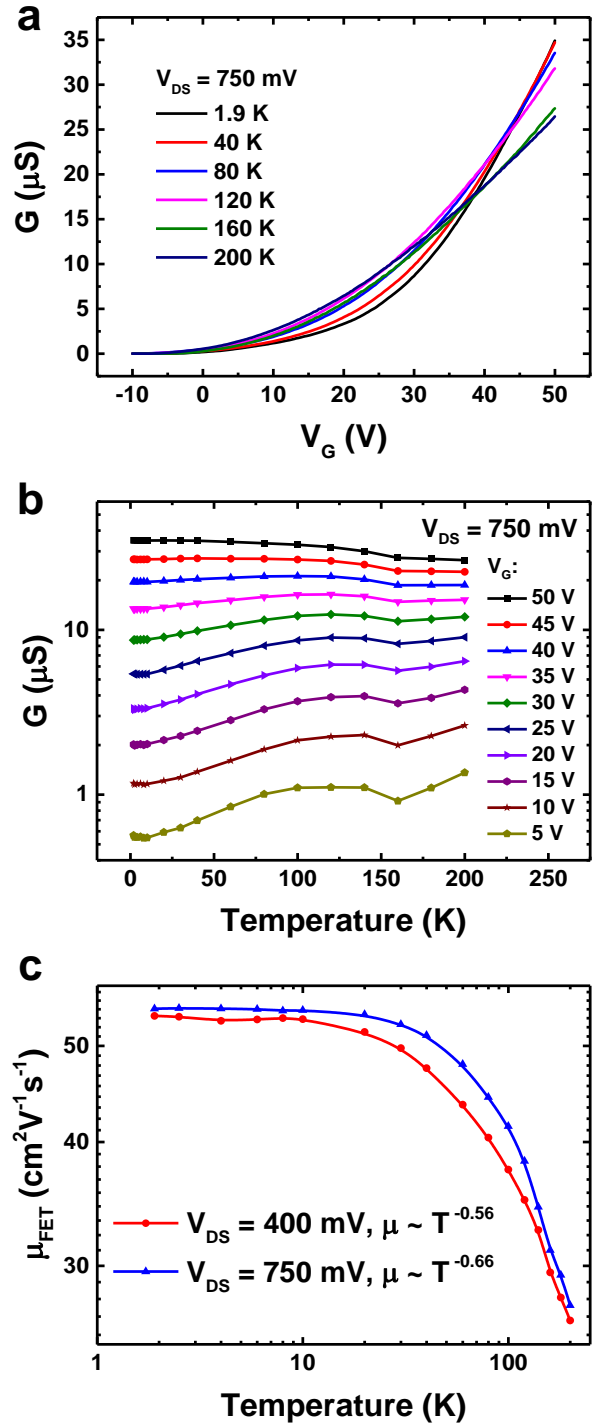

**Figure 3.** (a) Conductance *G* as a function of $V_G$ at different temperatures for the tri-layer $MoS_2$ device. (b) Temperature dependence of *G* at different $V_G$. When $V_G <$ 35 V the *G-T* curve shows an insulating behavior while $V_G >$ 35 V it becomes metallic. (c) Mobility as function of temperature under different $V_{DS}$. $\mu_{FET}$ is independent of *T* when *T* < 10 K, while above ~30 K $\mu_{FET}$ decreases following a $T^{-\gamma}$ dependence with $\gamma = 0.56 \sim 0.66$.

**Figure 4**

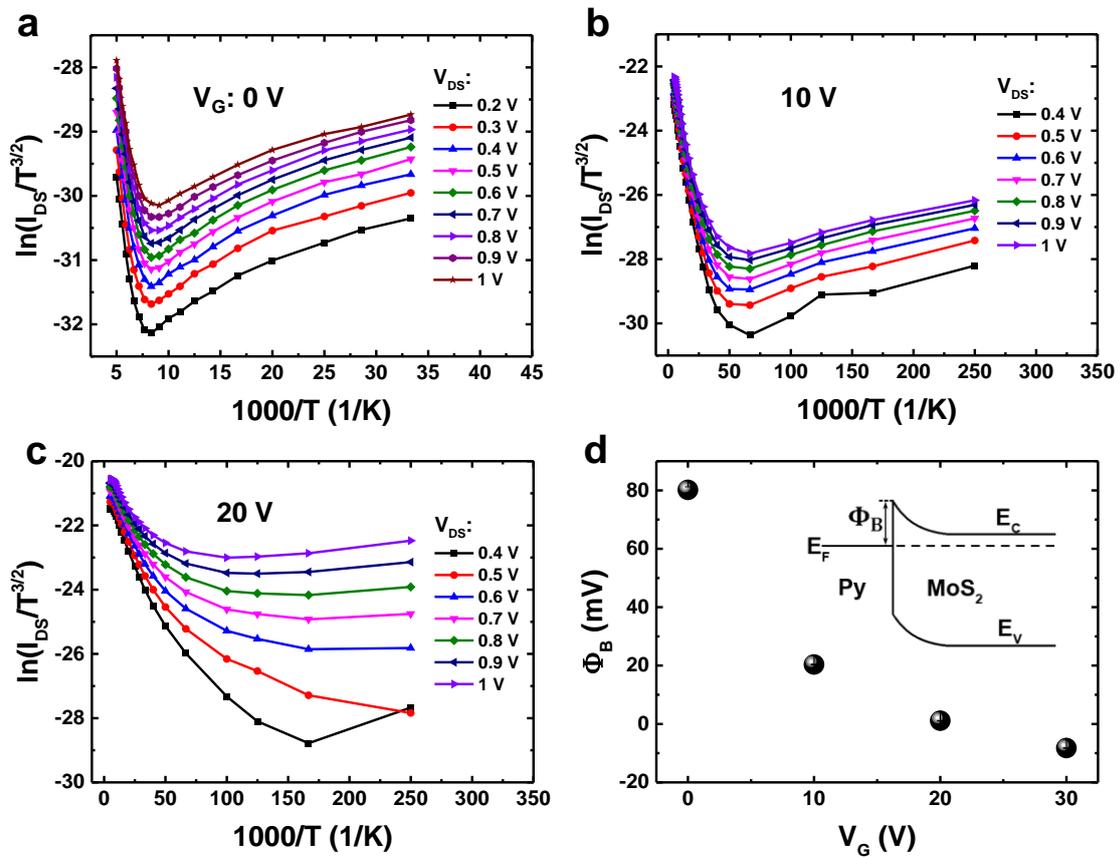

**Figure 4.** (a-c) Arrhenius plot of monolayer MoS$_2$ with Py electrodes in a large temperature range. A back gate voltage of $V_G$ = 0 V, 10, and 20 V was applied to the device as shown in (a), (b) and (c), respectively. (d) Schottky barrier height $\Phi_B$ as a function of $V_G$. The insets show the band diagram of Schottky barrier between Py and monolayer MoS$_2$.

**Figure 5**

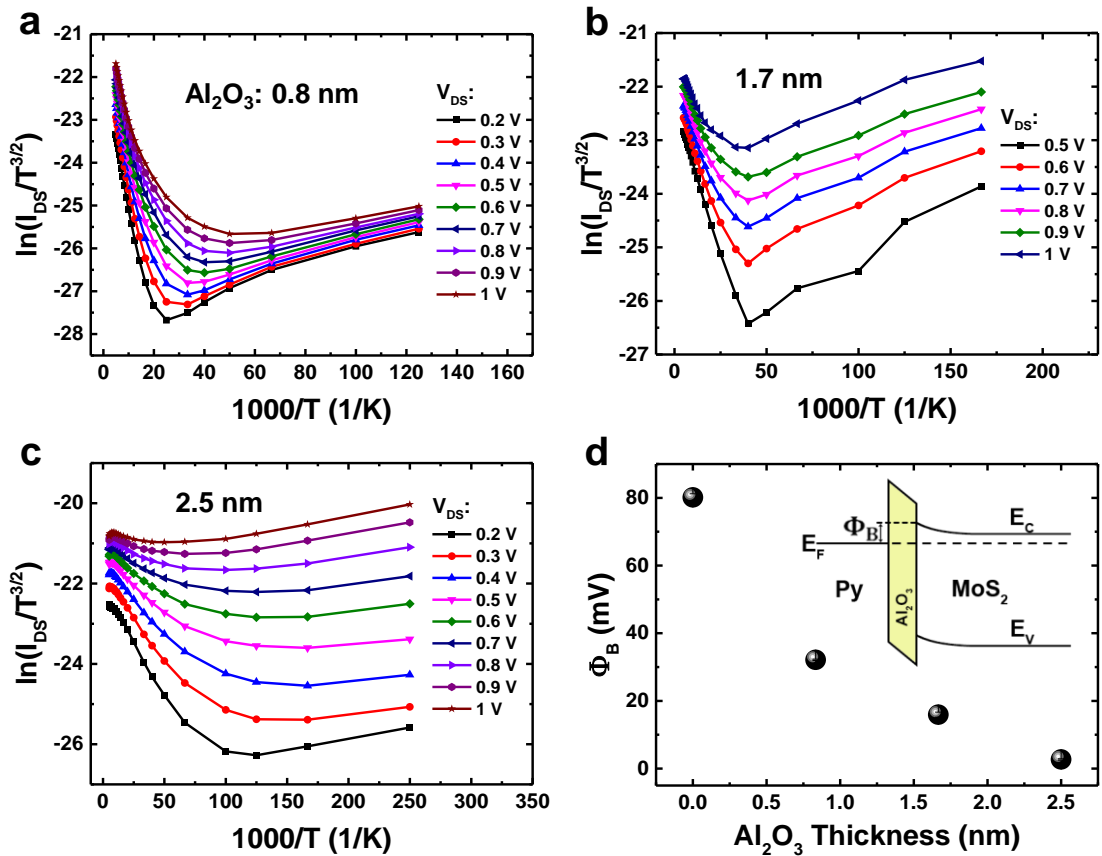

**Figure 5.** (a-c) Arrhenius plot of monolayer MoS₂ with different thickness of Al₂O₃ inserted. Al₂O₃ thickness is 0.8, 1.7 and 2.5 nm in (a), (b), and (c), respectively. (d) Schottky barrier height $\Phi_B$ as a function of the Al₂O₃ thickness. The inset shows the band diagram of Schottky barrier when inserting the Al₂O₃ tunneling layer.

**Acknowledgements**

This work was supported by the National Young 1000 Talent Plan, Pujiang Talent Plan in Shanghai, National Natural Science Foundation of China (61322407), and the Chinese National Science Fund for Talent Training in Basic Science (J1103204). Part of the sample fabrication was performed at Fudan Nano-fabrication Laboratory.

**Author contributions**

F.X. conceived the ideas and supervised the overall research. W.Y. fabricated the devices and carried out the entire characterizations. Y.W and T.T. contributed to the sample preparation and measurement. L.T. and Y.B. provided the monolayer $MoS_2$ and took the Raman spectrum. W.Y. and F.X. wrote the paper with help from all other co-authers.

**Additional information**

The authors declare no competing financial interests. Supplementary information accompanies this paper on www.nature.com/srep. Reprints and permissions information is available online at http://npg.nature.com/reprintsandpermissions. Correspondence and requests for materials should be addressed to F.X.